\documentclass[12pt]{article}
\usepackage{amsmath}
\usepackage{amsfonts}
\usepackage{amssymb}
\usepackage{latexsym}
\usepackage{graphicx}
\usepackage[english]{babel}

\begin{document}

\topmargin 0pt
\oddsidemargin 0mm



\begin{flushright}

\hfill{USTC-ICTS-08-04}\\

\end{flushright}
\vspace{4mm}

\begin{center}

{\Large \bf Non-Gaussianity, Isocurvature Perturbation,
  Gravitational Waves and a No-Go Theorem for Isocurvaton}

\vspace{8mm}

{\large Miao Li$^{1,2}$, Chunshan Lin$^1$, Tower Wang$^{2,1}$, Yi Wang$^{2,1}$}

\vspace{4mm}

{\em
$^{1}$ Interdisciplinary Center of Theoretical Studies, USTC,\\
Hefei, Anhui 230026, P.R.China\\
$^{2}$ Institute of Theoretical Physics, CAS,
Beijing 100080, P.R.China}

\end{center}

\vspace{6mm}

We investigate the isocurvaton model, in which the isocurvature
perturbation plays a role in suppressing the curvature perturbation,
and large non-Gaussianity and gravitational waves can be produced
with no isocurvature perturbation for dark matter. We show that in
the slow roll non-interacting multi-field theory, the isocurvaton
mechanism can not be realized. This result can also be generalized
to most of the studied models with generalized kinetic terms. We
also study the implications for the curvaton model. We show that
there is a combined constraint for curvaton on non-Gaussianity,
gravitational waves and isocurvature perturbation. The technique
used in this paper can also help to simplify some calculations in
the mixed inflaton and curvaton models. We also investigate
possibilities to produce large negative non-Gaussianity and nonlocal
non-Gaussianity in the curvaton model.

\newpage


\section{Introduction}

Inflation has been remarkably successful in solving some puzzles in
the standard hot big bang cosmology
\cite{Guth81,Linde82,Steinhardt82,Starobinsky-inf}. Inflation also
predicts that fluctuations of quantum origin were generated and
frozen to seed wrinkles in the cosmic microwave background (CMB)
\cite{CMBobserve,WMAP} and today's large scale structure
\cite{Mukhanov81,Guth82,Hawking82,Starobinsky82,Bardeen83}.

In spite of the success, inflation also faces some naturalness
problems. One of the problems is why the inflaton potential is so
flat, leading to typically $10^5$ e-folds instead of $50\sim 60$,
which is needed to solve the flatness and horizon problems.

Since the invention of inflation, a great number of inflation models were
proposed. Selecting the correct inflation model has become one of
the key problems in cosmology. The currently observed quantities
such as the power spectrum and the spectral index are not adequate
to distinguish the inflation models. However, luckily some more
quantities are expected to be measured accurately in the forthcoming
experiments. For example, non-Gaussianity, isocurvature
perturbation, and primordial gravitational waves.

Non-Gaussianity characterizes the departure of perturbations from
the Gaussian distribution. To characterize this departure, the non-Gaussian
estimator $f_{NL}$ is often used. Using the WMAP convention,
$f_{NL}$ can be written as \cite{fnl}
\begin{equation}
  \label{eq:fnl}
  \zeta=\zeta_g +\frac{3}{5} f_{NL}\zeta_g^2 ~,
\end{equation}
where $\zeta$ is the curvature perturbation in the uniform density
slice, and $\zeta_g$ is the Gaussian part of $\zeta$. This
particular model of non-Gaussianity is called the local shape
non-Gaussianity. The simplest single field inflation models predict
that $f_{NL}<{\cal
  O}(1)$. So large $f_{NL}$ indicates a departure from these
simplest models, such as the curvaton \cite{Lyth:2001nq} and (non-inflationary)
ekyrotic \cite{ekpyrotic}
models. To
describe more general shapes, such as the k-type\cite{K} and the DBI shapes \cite{DBI}, one needs to
calculate the 3-point correlation functions.

Recently, there are hints from experiments that the local shape
non-Gaussianity may be large. \cite{Yadav:2007yy} claims that
$f_{NL}=0$ is excluded above $99.5\%$ confidence level. In the WMAP
5-year data analysis, it is shown that the expectation value of
$f_{NL}$ using the bi-spectrum method is $f_{NL}=51$, but
$f_{NL}=0$ still lies within the $2 \sigma$ range. If the
non-Gaussianity is confirmed in the WMAP 8-year or the Planck
experiment, it will be a very powerful tool to distinguish between
inflation models.

Another possibility beyond the simplest single field inflation model is
isocurvature perturbation. The existence of isocurvature
perturbation indicates that there are more than one scalar degrees
of freedom during inflation. This can arise from multi-field
inflation \cite{Gordon:2000hv, Seery:2005gb}, modified gravity
\cite{Chen:2006wn}, or some exotic matter content during inflation
\cite{Chen:2006qy}. The WMAP 5-year +BAO+SN bound on
isocurvature perturbation is $\alpha_{-1}<0.0037$ (95\% CL). So no
evidence for isocurvature perturbation is shown. This result can be
used to constrain models such as the curvaton model.

The primordial gravitational waves also provide an important probe
for the early universe. The amplitude of gravitational waves varies
greatly in different inflation models. For example, chaotic
inflation predicts a tensor-to-scalar ratio $r\sim {\cal O}(0.1)$.
While most known stringy inflation models predict $r\lesssim {\cal
O} (10^{-3})$. The WMAP 5-year result, combined with BAO and SN
gives $r<0.2$ at $95\%$ confidence level. This has put a tight
constraint on chaotic inflation models. On the other hand, if future
experiments show that $r>{\cal O} (10^{-3})$, it will be a challenge
for string cosmology.

One attempt to solve the flat potential problem of inflation, and to
produce a large tensor-to-scalar ratio is proposed in
\cite{isocurvaton}. The idea is to suppress the perturbation outside
the inflationary horizon. This scenario looks like the curvaton
scenario, while the second scalar field only creates isocurvature
perturbation. So we call this scenario ``isocurvaton''. In this
paper, we show that if the isocurvaton scenario can be realized,
large non-Gaussianity can be produced, without producing observable
isocurvature perturbations.

However, it is not easy to realize the isocurvaton scenario. In
\cite{isocurvaton}, it is shown that in the simplest slow roll
inflation model, with $m^2\chi^2$ type isocurvaton, the above
scenario does not work. In \cite{Linde:2005he} it is argued that
the scenario does not work either for more general cases. In this
paper, we prove a no-go theorem that during slow roll inflation, if
the isocurvaton does not interact with the inflaton, then the
super-horizon perturbation can not be suppressed. The proof also
goes through for the k-type \cite{K} isocurvaton, where the
kinetic term of the isocurvaton and inflaton are allowed to be
generalized. After proving the no-go theorem, we discuss some
possible ways to bypass the theorem.

The proof of the no-go theorem also provides some new insights into
the combined inflaton and curvaton fluctuation \cite{Langlois:2004nn}. It is shown that in
the uniform inflaton density slice, the curvaton propagates
freely after the quantum initial condition is provided. This provides
a simplified treatment for the combined inflaton and curvaton
fluctuation.

We also combine the results of non-Gaussianity, isocurvature
perturbation and gravitational waves to constrain the curvaton
model. It is shown that the inequality $\label{f5} f_{NL}<
\frac{5}{432} r_T \left(\frac{M_p}{T}\right)^{2/3}$ should
be satisfied for the non-Gaussianity, gravitational waves and the
temperature of the universe when CDM was created.

This paper is organized as follows, in Section 2, we show the
virtues of the isocurvaton model, this is why the model is worth to
investigate. In Section 3, we prove a no-go theorem for the
isocurvaton scenario in the slow roll non-interacting multi-field
models. In Section 4, we extend the proof to the case with
generalized kinetic terms. In Section 5, we discuss the implications
for the curvaton model. In Section 6, we provide a simplified
analysis for the combined inflation and curvaton perturbations. In
Section 7, we discuss some other possibilities including large
negative $f_{NL}$ and large equilateral non-Gaussianity. We conclude
and discuss some possibilities to bypass the no-go theorem in
Section 8.

\section{Virtues of Isocurvaton}

In this section, we discuss the virtues of isocurvaton. We show that
some features of this scenario are rather attractive. That is why a
no-go theorem is valuable to mark the forbidden regions.

We use $\varphi$ to denote the inflaton
and use $\chi$ to denote the isocurvaton. It is shown in
\cite{Lyth:2004gb} that if there is no interaction between these two
components, the curvature perturbations for inflaton and isocurvaton on
their uniform density slices are separately conserved. The proof is
reviewed briefly in the appendix. These curvature
perturbations can be written in a gauge invariant form as
\begin{equation}
  \label{eq:zetas}
  \zeta_\varphi=-\psi-H\frac{\delta\rho_\varphi}{\dot\rho_\varphi}~,~~~
\zeta_\chi=-\psi-H\frac{\delta\rho_\chi}{\dot\rho_\chi}~,
\end{equation}
where $\psi$ is the metric perturbation. The explicit definitions for
$\psi$, $\delta\rho_\varphi$ and $\delta\rho_\chi$ are given in the
appendix. The total curvature perturbation takes the form
\begin{equation}
  \label{eq:zeta}
  \zeta = -\psi-H\frac{\delta\rho}{\dot\rho}= r
  \zeta_\chi+(1-r)\zeta_\varphi~,~~~ r\equiv \frac{\dot\rho_\chi}{\dot\rho_\chi+\dot\rho_\varphi}~.
\end{equation}

Note that if the inflaton and the isocurvaton have different equations
of state, $r$ will vary with respect to time. In this case, $\zeta$
is not conserved. Especially, during the epoch that inflaton has
decayed to radiation and the isocurvaton oscillates around its
minimum, we have $\rho_\varphi\propto a^{-4}$ and $\rho_\chi\propto
a^{-3}$. In this case, $r$ increases with time until the isocurvaton
decays. After curvaton decays to radiation, $r$ is a constant, and
$\zeta$ is conserved. If we would further assume $r\zeta_\chi\ll
(1-r)\zeta_\varphi$ when the isocurvaton decays, then

\begin{equation}
\label{eq:isocurvaton}
  \zeta=(1-r)\zeta_\varphi~.
\end{equation}
From (\ref{eq:isocurvaton}) we observe that if the isocurvaton decays
 very late so that $1-r\ll 1$, then the super Hubble horizon
perturbation is suppressed.

The direct consequence of suppressing the super-horizon perturbation
is to provide a solution to the problem of the flatness of the potential.
This can make inflation more natural, this is because a large scalar
type perturbation usually implies a non-flat potential.

The isocurvaton also serves as an amplifier for non-Gaussianity, this
is because if the initial inflaton fluctuation is larger, it
should generate a larger non-Gaussianity than the standard scenario.
This can be seen explicitly by writing
\begin{equation}
  \label{eq:fnlvarphi}
  \zeta_\varphi=\zeta_{\varphi g}+\frac{3}{5}f_{NL\varphi} \zeta_{\varphi g}^2
\end{equation}
Combining (\ref{eq:fnl}), (\ref{eq:isocurvaton}) and
(\ref{eq:fnlvarphi}), for the observable non-Gaussianity, we get
\begin{equation}
  \label{eq:fnlfinal}
  f_{NL}=\frac{1}{1-r} f_{NL\varphi}~.
\end{equation}
When $1-r\ll 1$, $f_{NL}$ can be large in the isocurvaton model.

For more general shape of non-Gaussianity, the 3-point function of
$\varphi$ is also amplified by isocurvaton. To see this, note that
\begin{equation}
  \label{eq:generalshape}
  \langle\zeta_{{\bf k}1}\zeta_{{\bf k}2}\zeta_{{\bf k}3}\rangle=
  (1-r)^3 \langle\zeta_{\varphi {\bf k}1}\zeta_{\varphi {\bf
    k}2}\zeta_{\varphi {\bf k}3}\rangle ~.
\end{equation}

Generally, one can rewrite the above 3-point functions using the
bi-spectrum expression
\begin{equation}
  \label{eq:bispectrum}
  \langle\zeta_{{\bf k}_1}\zeta_{{\bf k}_2}\zeta_{{\bf k}_3}\rangle
\propto \delta^3({\bf k}_1+{\bf k}_2+{\bf k}_3) {\cal P}_\zeta^2
f_{NL}^{\rm (nonlocal)}  {\cal A}(k_1,k_2,k_3)
  ~,
\end{equation}
where ${\cal P}_\zeta$ is the dimensionless power spectrum of
$\zeta$, and ${\cal A}(k_1,k_2,k_3)$ describes the shape of the
non-Gaussianity.

So for general shape non-Gaussianity we have
\begin{equation}
  \label{eq:nonlocal}
  f_{NL}^{\rm (nonlocal)}=\frac{1}{1-r} f_{NL\varphi}^{\rm (nonlocal)}~.
\end{equation}

As there are two components in the model, it is natural to ask
whether the isocurvature perturbation is produced in the model. The
treatment for isocurvature perturbation is the same as for the curvaton
model. If dark matter is produced after the isocurvaton decays, the
model can be consistent with the experimental results on
isocurvature perturbation.

In the isocurvaton scenario, the scalar perturbation is suppressed,
however, the tensor perturbation is unaffected. This results in an
enhancement for the tensor-to-scalar ratio. In this scenario,
the observed tensor-to-scalar ratio becomes
\begin{equation}
  r_T\equiv
  \frac{P_T}{P_\zeta}=\frac{1}{1-r}\frac{P_T}{P_{\zeta_\varphi}}=\frac{r_{T0}}{1-r}~,
\end{equation}
where $r_{T0}$ is the tensor-to-scalar ratio without the isocurvaton
dilution. If the isocurvaton scenario works, and future
experiments detect gravitation waves, then isocurvaton with $1-r\ll
1$ can be a way to save a large number of string inflation models,
which make the prediction that the gravitational waves are too small
to be detected. This possibility is investigated in detail in
\cite{isocurvaton}.

However, unluckily, as we shall show in the following two sections,
under reasonable assumptions, the isocurvaton scenario can not be
realized.

\section{A No-Go Theorem for Isocurvaton}

As stated in the introduction, there are theoretical obstructions to
constructing the isocurvaton model. In this section, we prove a
no-go theorem that the isocurvaton model can not be realized in the
standard non-interacting slow roll double field models.

We shall prove that in the $\delta\rho_\varphi=0$ gauge, to good
approximation, $\delta\chi$ propagates freely, and do not feel the
inflaton fluctuation or the gravitational potential. With this
result, the gauge invariant curvature perturbation can be written as
\begin{equation}
  \zeta_\varphi=-\psi~,~~~\zeta_\chi=-\psi-H\frac{\delta\rho_\chi}{\dot\rho_\chi}=-\psi+\frac{\dot{\delta\chi}}{3\dot\chi}+\frac{1}{3}\frac{V_\chi}{\dot\chi^2}\delta\chi~,
\end{equation}
where as we shall prove, $\delta\chi$ is an independent stochastic
source other than $\psi$. So the $-\psi$ term in $\zeta_\chi$ can
not be canceled without fine-tuning , and $\zeta_\chi \ll
\zeta_\varphi$ can not be naturally realized. This result indicates
that the model considered above can not realize the isocurvaton
scenario.

To prove the free propagation of $\delta\chi$ in the
$\delta\rho_\varphi=0$ gauge, we first show that outside the
horizon, $\delta\chi$ propagates freely without gravitational source
term. After that, we show that the initial condition for $\delta
\chi$ is determined by the quantum fluctuation before horizon exit,
and the influence from the inflaton fluctuation and the
gravitational potential can be neglected.

We start with the familiar Newtonian gauge perturbation
equations. Before curvaton dominates the energy density, the
perturbation equations takes the form
\begin{equation}\label{n00}
  -3H (H\psi^{\rm (n)}+\dot\psi^{\rm (n)})-\frac{k^2}{a^2}\psi^{\rm
    (n)}=4\pi G \delta\rho_\varphi^{\rm (n)}~,
\end{equation}
\begin{equation}\label{n11}
  (2\dot H+3H^2)\psi^{\rm (n)}+ 4H\dot\psi^{\rm (n)}+\ddot\psi^{\rm
    (n)}=4\pi G\delta p_\varphi^{\rm (n)}~,
\end{equation}
\begin{equation}\label{nchi}
  \ddot{\delta\chi}^{\rm (n)}+ 3H \dot{\delta\chi}^{\rm (n)}
+V_{\chi\chi}\delta\chi^{\rm (n)}=-2V_\chi\psi^{\rm (n)}+4\dot\chi\dot\psi^{\rm (n)}~,
\end{equation}
where the superscript ``(n)'' denotes the Newtonian gauge. The gauge
transformation from the Newtonian gauge to the $\delta\rho_\varphi=0$
gauge can be written as
\begin{equation}\label{gauge}
  \psi^{\rm (n)}=\psi-H\beta~,~~~\delta x^{\rm (n)}=\delta x+\dot x
  \beta~,~~~
\beta\equiv \frac{\delta\rho_\varphi^{\rm (n)}}{\dot\rho_\varphi}~,
\end{equation}
where $x=x(t)$ denotes a background scalar field, and $\delta x$
stands for its perturbation. We assume that
$p_\varphi=p_\varphi(\rho_\varphi)$, so that in the
$\delta\rho_\varphi=0$ gauge, we have $\delta p_\varphi=0$. This
assumption holds for the ideal fluid without intrinsic isocurvature
perturbation, as well as the scalar field outside the inflationary
horizon. The proof for the ideal fluid is straightforward, and the
proof for the scalar field is given in the appendix.

We first consider the $k\ll aH$ limit.
Changing the equations into the $\delta\rho_\varphi=0$
gauge, one can simplify Eqs. (\ref{n00}) and (\ref{n11}) as
\begin{equation}\label{0011}
  \dot\psi=0~,~~~\dot\beta=\psi-H\beta~,
\end{equation}
note that in this paper, all perturbation variables without the
superscript (n) denote perturbations in the $\delta\rho_\varphi=0$
gauge if not stated otherwise.

Writing Eq. (\ref{nchi}) in the $\delta\rho_\varphi=0$ gauge, and
using Eq. (\ref{0011}), we find
\begin{equation}
  \label{eq:freechi}
  \ddot{\delta\chi}+3H\dot{\delta\chi}+V_{\chi\chi}\delta\chi=0~.
\end{equation}
The $\psi$ and $\beta$ terms are canceled in this equation. In other
words, in this gauge,  $\chi$ does not feel the gravitational
potential and propagates freely.

This result can be obtained in a simpler way. One can show
that $\delta\chi$ is proportional to the isocurvature perturbation.
Then from the well-known result in double field inflation that
isocurvature perturbation propagates without source outside the
horizon, we obtain that $\delta\chi$ propagates freely. However, we
still write down the derivation explicitly, because this derivation
is rather general, holds after $\varphi$ decays, and can be used to
simplify some calculations in the curvaton model.

Now let us consider the $k\gg aH$ and $k\sim aH$ case, and see whether
$\delta\chi$ can feel the gravitational potential. Note that the super
horizon analysis only requires $p_\varphi=p_\varphi(\rho_\varphi)$,
and does not require detailed information about the
inflaton. While to investigate the horizon crossing, we need to
focus on the standard single field inflaton plus the isocurvaton.

We employ the results in the double field inflation model
\cite{Gordon:2000hv} to rewrite $\varphi$ and $\chi$ into
the inflation direction $\sigma$ and the isocurvature direction $s$,
\begin{equation}
  \label{eq:sigmas}
  \delta\sigma\equiv\cos\theta\delta\varphi+\sin\theta\delta\chi~,~~~
\delta s\equiv -\sin\theta\delta\varphi+\cos\theta\delta\chi~,~~~
\sin\theta\equiv\frac{\dot\chi}{\sqrt{\dot\varphi^2+\dot\chi^2}}~.
\end{equation}
Note that $\delta s$ is automatically gauge invariant. During
inflation, if $\dot\chi\ll \dot\varphi$, we have $\theta\simeq 0$
during inflation. The inflation direction does not change and
the isocurvature perturbation is obviously sourceless. However, we
do not limit to this case, because we only require
$\rho_\chi\ll\rho_\varphi$ during inflation.

The perturbation equation for the isocurvature direction can be written as
\begin{equation}\label{ds}
  \ddot{\delta s}+3H\dot{\delta
    s}+\left(\frac{k^2}{a^2}+V_{ss}+3\dot\theta^2\right)\delta s=
\frac{\dot\theta}{\dot\sigma}\frac{k^2}{2\pi G a^2}\psi^{\rm (n)}~.
\end{equation}
Now we shall prove that the RHS of Eq. (\ref{ds}) is much smaller
than a typical term in the LHS. To see this, we first estimate the
fluctuation amplitude of the inflation direction. From the
perturbation in the Newtonian gauge $\dot\psi^{\rm (n)}+H\psi^{\rm
(n)}=4\pi G\dot\sigma\delta\sigma^{\rm (n)}$ and the slow roll
condition, we have
\begin{equation}
  \left|\frac{\dot\sigma}{H}\psi^{\rm (n)}\right|\leq \left|\frac{4\pi G\dot\sigma^2\delta\sigma^{\rm
  (n)}}{H^2}\right| \ll \delta\sigma^{\rm (n)}~.
\end{equation}
From the amplitude for $\delta\sigma$ in the $\psi=0$ gauge
\cite{Gordon:2000hv}, we have
\begin{equation}
  \delta\sigma^{\rm (n)}\simeq \delta\sigma^{\rm (n)}+\frac{\dot\sigma}{H}\psi^{\rm (n)}= \left(\delta\sigma\right)_{\psi=0~\rm
  gauge}\sim a^{-1}k^{-1/2} e^{-ik\tau}~,
\end{equation}
Next, from the perturbation equation $\dot\psi^{\rm (n)}+H\psi^{\rm
(n)}=4\pi G\dot\sigma\delta\sigma^{\rm (n)}$, we have
\begin{equation}
  |\psi^{\rm (n)}|\sim k^{-3/2}\frac{\dot\sigma}{M_p^2}~,
\end{equation}
where we only want to count the orders in slow roll parameters, so the
numerical coefficients are neglected.
The source term in the RHS
of (\ref{ds}) takes the form
\begin{equation}
  \left|\frac{\dot\theta}{\dot\sigma}\frac{k^2}{2\pi G a^2}\psi^{\rm (n)}\right|\sim
\left|\frac{\dot\chi\ddot\varphi-\ddot\chi\dot\varphi}{\dot\varphi^2+\dot\chi^2}\right|
\times\frac{k^2}{a^2} k^{-3/2}\ll H \frac{k^2}{a^2} k^{-3/2}~,
\end{equation}
where we have used the slow roll approximation
\begin{equation}
  \left|\frac{\dot\chi\ddot\varphi-\ddot\chi\dot\varphi}{\dot\varphi^2+\dot\chi^2}\right|
\leq\left|\frac{\dot\chi\ddot\varphi-\ddot\chi\dot\varphi}{2\dot\varphi\dot\chi}\right|\leq\left|\frac{\ddot\varphi}{2\dot\varphi}\right|+\left|\frac{\ddot\chi}{2\dot\chi}\right|\ll H
\end{equation}
However the quantum initial condition of $\delta s$ is $|\delta s| \sim
a^{-1}k^{-1/2}$, so when $k\geq aH$, for a typical term in the LHS of (\ref{ds}),
\begin{equation}
  \left|\frac{k^2}{a^2}\delta s\right| \geq H \frac{k^2}{a^2} k^{-3/2} \gg
  \left|\frac{\dot\theta}{\dot\sigma}\frac{k^2}{2\pi G
      a^2}\psi^{\rm (n)}\right|~.
\end{equation}
Since the horizon exit does not take too many e-folds, the
suppression in slow-roll parameter indicates that the initial
condition of $\delta s$ is prepared by the quantum fluctuation, and
the influence from the gravitational potential can be ignored.
Physically, this result originates from the fact that the inflaton
and the isocurvaton fields couple weakly due to the slow roll
conditions.

Note that we have ignored the back-reaction of $\delta s$ on the
inflaton direction. This approximation can also be verified using the
slow roll approximation.

After horizon exit, in the $\delta\rho_\varphi=0$ gauge,
$\delta\varphi\simeq0$ \cite{Lyth:2004gb}. So in this case, $\delta
s\simeq\cos\theta\delta\chi$. Combining the above discussion for
$\delta s$, we conclude that the initial condition for $\delta\chi$
is prepared by its quantum fluctuation.

Finally, recall (\ref{eq:freechi}), we get to the conclusion that
$\delta\chi$ has its independent quantum initial condition,
evolves freely and does not feel the gravitational potential.


This proof of the no-go theorem can be generalized directly to the
non-interacting multi-field isocurvaton case. So increasing the number
of fields does not make things better.

There are two exceptions where the above proof does not apply,
namely, the vacuum energy and a field with completely flat potential
as the isocurvaton. However, neither of them can serve as
isocurvaton. For the vacuum energy, we can think of it as a shift of
the inflaton potential, so it can not dilute the inflaton
perturbation. For a field with completely flat potential, the
solutions for both $\sigma$ and $\delta\sigma$ have a constant mode
plus a decaying mode. The constant mode does not contribute to
$\zeta_\sigma$. So up to the decaying mode, the flat potential case
is the same as the vacuum energy case, and can not serve as
isocurvaton.

In this section, we have provided a direct and self-contained proof
for the no-go theorem on isocurvaton. In order to link the double
field inflaton to observations, analysis similar to the above proof
has been performed in the literature. In a series of papers
\cite{doublefield}, the authors proved that the cross correlation
between the adiabatic and entropy modes is suppressed by the slow
roll parameters, and the primordial adiabatic mode is related to the
adiabatic mode at horizon crossing by
\begin{equation}
P_\zeta=\frac{P_{\zeta*}}{\sin^2\Theta}~,~~~\sin\Theta\equiv\frac{1}{\sqrt{1+{\cal
T}_{\cal RS}^2}}~,
\end{equation}
where ${\cal T}_{\cal RS}$ is the transfer function from entropy
mode to adiabatic mode. From this relation, one can see that the
super horizon perturbation can not be suppressed in the context of
slow roll inflation. This reasoning also extends to the interacting
double field theory with some additional slow roll assumptions.

\section{Proof for Generalized Kinetic Terms}

In this section, we try to generalize the no-go theorem in the last
section to the case of generalized kinetic terms. As done in the
last section, we first investigate the super horizon evolution, and
then study the horizon crossing.

Consider the isocurvaton Lagrangian $P=P(X(\chi),\chi)$, where
$X(\chi)=\frac{1}{2}g^{\mu\nu}\partial_\mu\chi\partial_\nu\chi$, and
a general dominate component originating from the inflaton
$p_\varphi=p_\varphi(\rho_\varphi)$. In the $k\ll aH$ limit, the
coupling equations for $\rho_\varphi$, $p_\varphi$ and $\phi$
(\ref{n00}), (\ref{n11}) are not changed, so we still have
(\ref{0011}).

The equation of motion for $\chi$ takes the form
\begin{equation}
  P_X g^{\mu\nu}\nabla_\mu\nabla_\nu\chi+\partial_\mu P_X
  g^{\mu\nu}\partial_\nu \chi-P_\chi=0 ~.
\end{equation}
Expanding this equation to the zeroth and first order in the
perturbation variables, we get the background and the leading order
perturbation equations in the Newtonian gauge,
\begin{equation}\label{generalizedbackground}
  \partial_t (P_X\dot\chi) + 3HP_X\dot\chi-P_\chi=0~,
\end{equation}
\begin{equation}\label{generalizedpert}
  \ddot\chi\delta P_X^{\rm (n)} + P_x\ddot{\delta\chi}^{\rm
  (n)}+\dot\chi\dot{\delta P}_X^{\rm (n)}+\dot P_X\dot{\delta\chi}^{\rm
  (n)} + 3H\dot\chi\delta P_X^{\rm (n)}+3HP_X\dot{\delta\chi}^{\rm
  (n)} - 2P_\chi\phi^{\rm (n)}-4 P_X\dot\chi\dot\phi^{\rm
  (n)}-\delta P_\chi^{\rm (n)}=0~.
\end{equation}

Terms such as $\delta P_\chi^{\rm (n)}$ in Eq.
(\ref{generalizedpert}) can be expanded into more explicit forms.
But we do not need this expansion for our purpose. Note that $X$,
$P_X$ and $P_\chi$ are scalars under the gauge transformation
(\ref{gauge}). Using the background equation of motion
(\ref{generalizedbackground}), it can be shown that in the
$\delta\rho_\varphi=0$ gauge, all the source terms in
(\ref{generalizedpert}) vanish,
\begin{equation}\label{generalizedpert}
  \ddot\chi\delta P_X + P_x\ddot{\delta\chi}+\dot\chi\dot{\delta P}_X
  +\dot P_X\dot{\delta\chi} + 3H\dot\chi\delta P_X+3HP_X\dot{\delta\chi}
  -\delta P_\chi=0~.
\end{equation}

Again, this result is not surprising, as isocurvature
perturbation should be sourceless after horizon crossing.

For horizon crossing, we focus on the model that the inflaton and
the isocurvaton have a unified generalized kinetic term. The
Lagrangian of the model takes the form
\begin{equation}
  P=P(X,\varphi,\chi)~,~~~
  X=\frac{1}{2}G_{IJ}\nabla_{\mu}\varphi^I\nabla^\mu\varphi^J~,
\end{equation}
where $G_{IJ}$ is the metric in the field space, and $I,J=\{1,2\}$
such that $\varphi^1=\varphi$, $\varphi^2=\chi$.

Using the results obtained in \cite{Langlois:2008mn}, the
isocurvature direction perturbation (\ref{eq:sigmas}) can be written
as
\begin{equation}
\ddot{\delta s}+\left(3H+\frac{\dot P_X}{P_X}\right)\dot{\delta s}+
\left(\frac{k^2}{a^2}+\mu_s^2+\frac{\Xi^2}{c_s^2}\right)\delta s
=-\frac{\dot\sigma}{\dot H} \Xi \frac{k^2}{a^2}\psi^{\rm (n)}~.
\end{equation}
where
\begin{equation}
  \Xi \equiv \frac{1}{\dot\sigma
  P_X}\left((1+c_s^2)P_s-c_s^2P_{Xs}\dot\sigma^2\right)~,
\end{equation} and
\begin{equation}
  \mu_s^2\equiv -\frac{P_{ss}}{P_X}+\frac{1}{2}\dot\sigma^2\bar R
  -\frac{1}{2 c_s^2 X}\frac{P_s^2}{P_X^2}+2\frac{P_{Xs}P_s}{P_X^2}~,
  ~~~c_s\equiv \frac{P_X}{P_X+2X P_{XX}}~,
\end{equation}
where $\bar R$ is the scalar curvature in the field space. Note that
$\Xi$ plays the role of $\theta$ in the standard kinetic term case,
which characterizes the coupling between the inflaton direction and
the isocurvature perturbations.

When $\Xi=0$, the isocurvaton propagates freely at horizon crossing.
So we still have that $\zeta_\chi$ equals $\zeta_\varphi$ plus an
term originating from an independent random quantum initial
fluctuation, and the isocurvaton scenario does not work.

If $\Xi$ is large enough to provide source to the isocurvature
perturbation, the above proof breaks down. However, the $\Xi\neq 0$
case has not been investigated analytically in the literature, with
only numerical results available (see, {\it e.g.}
\cite{Yang:2008ns}). We are not able to prove the no-go theorem in
this case.

Note that in the proof of the horizon crossing, the generalized
kinetic term will introduce interaction between $\varphi$ and
$\chi$. However, our proof makes no reference to the conserved
quantities during inflation. So the interaction is not an
obstruction of our proof.


\section{Implication for the Curvaton Models}

In this section, we first show that the no-go theorem proved above
does not rule out the curvaton scenario. We also discuss some
physical constraint for the curvaton model from the non-Gaussianity,
isocurvature perturbation and gravitational waves experimental data.

Let us see why the curvaton scenario is not affected by the no-go
theorem. Most of the calculation in the above two sections applies
to the curvaton scenario by setting $\psi^{\rm (n)}=0$. But in the
curvaton scenario, it is the curvaton field, not the inflaton field,
which produces the primordial perturbations. In the curvaton scenario,
$\zeta_\varphi$ is small, and does not need to be canceled by the
curvaton field. So the no-go theorem does no harm to the curvaton
scenario.

However, the observables we consider, namely, non-Gaussianity,
isocurvature perturbation and gravitational waves do put a tight
constraint on the curvaton scenario. In the remainder of this
section, we shall combine the ``non-Gaussianity + gravitational
waves'' \cite{Huang:2008ze} and the ``non-Gaussianity + isocurvature
perturbation'' constraints \cite{WMAP, Lyth:2002my, Beltran:2008ei} to provide a
more complete constraint for the curvaton model. 

We first quickly review the result in \cite{Huang:2008ze}. Consider
the simplest curvaton model. To distinguish it from the inflaton
direction used in the above sections, we denote the curvaton field
by $\chi$. The local shape non-Gaussianity is related to the ratio
of energy densities when curvaton decays
\begin{equation}
  f_{NL}\simeq \frac{5}{4r}~,~~~r=\left(\frac{\rho_\chi}{\rho_{\rm
  tot}}\right)_D~.
\end{equation}
Note that by writing this equation, we have assumed $f_{NL}>1$.
Otherwise, the order 1 and order $r$ terms in $f_{NL}$ can dominate
the expression.

The curvaton starts to oscillate after inflaton decays into
radiation. As the curvaton is much lighter than the inflation scale,
the field value of the curvaton field is practically unchanged from
the time of
horizon exit to the time the curvaton starts to oscillate. So when the curvaton starts to
oscillate, we have
\begin{equation}
  H=m~,~~~\rho_\chi=\frac{1}{2}m^2\chi_*^2~,~~~\rho_\varphi=3m^2M_p^2~,
\end{equation}
where $m$ is the curvaton mass, and $\chi_*$ is the curvaton field
value at horizon exit.

Another important time scale in the curvaton scenario is the time when the curvaton
decays. When the curvaton decays, we have
\begin{equation}
  H=\Gamma~, ~~~\rho_\varphi=3M_p^2\Gamma^2~.
\end{equation}

From $\rho_\chi\propto a^{-3}$ and $\rho_\varphi\propto a^{-4}$, we
have
$\rho_\chi=\frac{\chi_*^2}{6M_p^2}\left(m/\Gamma\right)^{1/2}\rho_\varphi$
when the curvaton decays. So
\begin{equation}
  r=\frac{\chi_*^2}{6M_p^2}\left(\frac{m}{\Gamma}\right)^{1/2}\end{equation}
  In terms of $f_{NL}$, we have
  \begin{equation}\label{f1}
  f_{NL}=\frac{15}{2}\frac{M_p^2}{\chi_*^2}\left(\frac{\Gamma}{m}\right)^{1/2}
\end{equation}
On the other hand, from the power spectrum
\begin{equation}\label{f2}
  P_\zeta^{1/2}=\frac{1}{3\pi} r \frac{H_*}{\chi_*}=\frac{5}{12\pi} \frac{1}{f_{NL}}
  \frac{H_*}{\chi_*}~.
\end{equation}
Use (\ref{f1}) and (\ref{f2}) to cancel the unknown $\chi_*$, we
have
\begin{equation}\label{f3}
  f_{NL}=\frac{5}{432} r_T
  \left(\frac{m}{\Gamma}\right)^{1/2}~,
\end{equation}
where $r_T\equiv {P_T}/{P_\zeta}$ is the tensor-to-scalar ratio, and
$P_T\equiv 2H_*^2/(\pi^2M_p^2)$ is the tensor mode power spectrum.

Finally, note that the curvaton decay rate $\Gamma$ should be larger
than the decay rate via gravitational coupling,
\begin{equation}\label{gammag}
  \Gamma>\Gamma_g\simeq\frac{m^3}{M_p^2}~.
\end{equation}

The above results in this section have been present in
\cite{Huang:2008ze}. Now we use them together with the isocurvature
constraint to derive a new inequality. Use (\ref{gammag}) to cancel
$m$ in (\ref{f3}), we have
\begin{equation}\label{f4}
  f_{NL}<\frac{5}{432}r_T\left(\frac{M_p}{\Gamma}\right)^{1/3}~.
\end{equation}

It is shown in \cite{WMAP, Lyth:2002my, Beltran:2008ei} that to avoid a large
isocurvature perturbation, cold dark matter (CDM) should be produced
after the curvaton decays. So we have
\begin{equation}
  \Gamma>H_{\rm CDM}~,
\end{equation}
where $H_{\rm CDM}$ denotes the Hubble parameter when CDM is
produced. $H_{\rm CDM}$ can be related with the temperature of the
universe $T$ when CDM decays as
\begin{equation}
  H_{\rm CDM}\simeq T^2/M_p~.
\end{equation}

Using Eq. (\ref{f4}), we have
\begin{equation}\label{f5}
f_{NL}<  \frac{5}{432} r_T \left(\frac{M_p}{T}\right)^{2/3}~.
\end{equation}

This bound should be used combined with the bound given in
\cite{Huang:2008ze},
\begin{equation}\label{qg}
  f_{NL}< 522 r_T^{1/4}~.
\end{equation}
The more constraining one should be used to get the final constraint.

\begin{figure}
    \center
    \includegraphics[width=11cm]{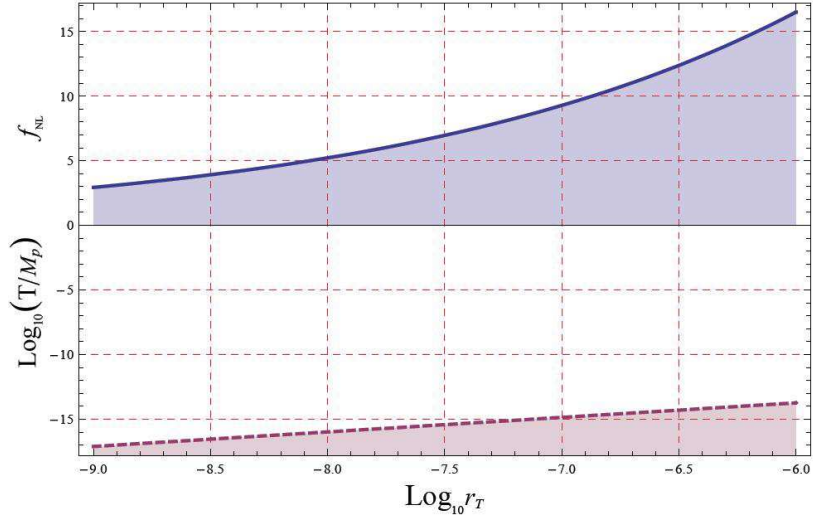}
    \caption{In this figure, we show the constraint (\ref{f5}) and (\ref{qg}).
    The up plate of the figure denotes the allowed region of $f_{NL}$ as
    a function of $r_T$ from (\ref{qg}). The shaded region is the allowed region.
    The down plate of the figure denotes the temperature when CDM decays which saturates the
    inequalities. In the shaded region, (\ref{qg}) is more strict than (\ref{f5}).
    In the unshaded region, (\ref{f5}) is more strict than (\ref{qg}).  }
\end{figure}

For example, assume $f_{NL}$ is produced by the curvaton scenario.
If $f_{NL}=50$, then $r_T>10^{-4}$. If the lower bound $r_T=10^{-4}$
is saturated, we get $T< 10^{-11} M_p\sim 10^{7}$GeV. This
constraint is not very tight for $m_{\rm CDM}$. However, it already
rules out some CDM candidates within the curvaton scenario, such as
invisible axions, magnetic monopoles and pyrgons.

On the other hand, if $f_{NL}= 5$ then $r_T>10^{-8}$. This seems
more natural in the small field inflation models. In this case, if
the lower bound is saturated, we get $T<10^{-17}M_p\sim 10$GeV. This
further rules out some dark matter candidates such as primordial black
holes.


The inequality (\ref{f5}) is applicable until $f_{NL}\sim 1$. After
that, other corrections for $f_{NL}$ begins to dominate. In this
limit, $r_T>10^{-11}$. If this bound is saturated, then
$T<10^{-19}M_p\sim 100$ MeV. This bound becomes borderline for LSP
and quark nuggets type dark matter.


Finally, we would like to compare our result with a recent paper
\cite{Beltran:2008ei}. In \cite{Beltran:2008ei}, the author also aims
to get a bound for the temperature of $f_{NL}$, $r_T$ and
$T$. The difference between our work and \cite{Beltran:2008ei} is that
we use different methods to constrain the curvaton mass
$m$. \cite{Beltran:2008ei} uses the spectral index, which gives
$m<0.1H_*$. While we use Eq. (\ref{gammag}) to constrain $m$. This
difference leads to different final results. In \cite{Beltran:2008ei},
the constraint takes the form
\begin{equation}\label{Beltran}
  T < 1.9\times 10^{-4} r_T^{5/4}f_{NL}^{-1}M_p~.
\end{equation}
When $r_T^{1/2}<0.023f_{NL}$, our bound Eq. (\ref{f5}) is more tight
than Eq. (\ref{Beltran}). When $r_T^{1/2}>0.023f_{NL}$,
Eq. (\ref{Beltran}) is more tight. 

\section{Improved Treatment for Combined Inflaton and Curvaton
  Perturbations}

Using the techniques developed in this paper, we are able to
simplify some calculations for the mixed inflaton and curvaton
scenario in the literature.

Note that the perturbations in the curvaton scenario (setting
$\psi^{\rm (n)}=0$), isocurvaton scenario, and mixed inflaton and
curvaton scenario are the same. The simplifications arises because
we have chosen the $\delta\rho_\varphi=0$ gauge. This gauge choice,
or rearranging the variables, can diagonalize the perturbation
equations outside the horizon.

Consider the scenario investigated in \cite{Langlois:2004nn}, with the curvaton
evolving in the radiation dominated era. Using the method in
\cite{Langlois:2004nn}, one is forced to solve the Bessel equation with
source term
\begin{equation}
  \ddot{\delta\chi}^{\rm (n)}+
  3H\dot{\delta\chi}^{\rm (n)}+m^2\delta\chi^{\rm (n)}=4\dot\chi\dot\psi^{\rm (n)}-2m^2\chi\psi^{\rm
  (n)}~, ~~~H=\frac{1}{2t}~.
\end{equation}
Although this equation can be solved analytically, it saves
some calculation and be easier to generalize if one rewrites
this equation in a sourceless manner. This simplification is just
what we have done in Eq. (\ref{eq:freechi}), where we have
considered a general potential for the curvaton and a general equation
of state for the inflaton.

For the $m^2\chi^2$ type curvaton potential, Eq. (15) reads
\begin{equation}
  \ddot{\delta\chi}+ 3H\dot{\delta\chi}+m^2\delta\chi=0~.
\end{equation}
Note that this equation has the same form with the background
evolution for $\chi$. So without solving any differential equations,
we know that for the non-decaying solution,
\begin{equation}
  \delta\chi=\frac{\chi}{\chi_0}\delta\chi_0~,
\end{equation}
where $\delta\chi_0$ is an integration constant. Use (\ref{gauge})
to translate this result into the Newtonian gauge, we have
\begin{equation}
  \delta\chi^{\rm (n)}=\frac{\chi}{\chi_0}\left(\delta\chi_0^{\rm (n)}
  -\dot\chi_0 t_0 \psi_0^{\rm (n)}\right)+t\dot\chi\psi^{\rm (n)}~.
\end{equation}
Note that $\delta\chi_0^{\rm (n)}
  -\dot\chi_0 t_0 \psi_0^{\rm (n)}$ is just a constant, which can
  be redefined to $\delta\chi_0^{\rm (n)}$. Then we recover one of the
  key results in \cite{Langlois:2004nn},
\begin{equation}
  \delta\chi^{\rm (n)}=\frac{\chi}{\chi_0}\delta\chi_0^{\rm (n)}
+t\dot\chi\psi^{\rm (n)}~,
\end{equation}
where we have not made any reference to the equation of state of
the inflaton component. Other results for the $m^2\chi^2$ potential
in \cite{Langlois:2004nn} can be recovered similarly.


\section{Generalizations for the Curvaton Model}

In this section, we investigate the possibility for large negative
$f_{NL}$ and nonlocal shape $f_{NL}$ in the curvaton model. This
possibility can be realized by phantom curvaton and k-curvaton
respectively. These models seem exotic, and not supported by any
evidence so far. For example, the vacuum stability and the
quantization problem in the phantom model are not solved. However,
we investigate these models as pure phenomenological possibilities.

In reference \cite{WMAP}, the WMAP5 data has
been analyzed by two different methods. It is intriguing that the two
methods prefer central values
of $f_{NL}^{local}$ opposite in sign. In particular, both at the
$95\%$ confidence level, the bispectrum analysis gives the best
estimate $-9<f_{NL}^{local}<111$, while the analysis of Minkowski
functionals prefers $-178<f_{NL}^{local}<64$ in contrast. It is
still unclear why they are so different. However, if one
naively disregards the bispectrum analysis for the moment, and takes
seriously the central value from Minkowski functionals, we will be
motivated to search for models with $f_{NL}^{local}\ll-1$. Let us
see what will happen if the curvaton component is phantom-like
\cite{Caldwell:1999ew}. In that case, the Eqs. (\ref{eq:zeta}) and
(\ref{eq:zetas}) are still valid. However, now $\dot\rho_\chi>0$. So
we have $r<0$, and $\zeta_\chi$ has the different sign from $\zeta$.
It is well-known that a different sign in $\zeta$ should produce a
different sign in $f_{NL}$. The calculation for $f_{NL}$ goes
through in the phantom model, so when $|r|\ll 1$, we have
\begin{equation}
  f_{NL}\sim \frac{1}{r}\ll -1~.
\end{equation}

Although the bispectrum analysis of WMAP5 (which is widely taken as
the best estimate) does not prefer a large negative $f_{NL}$, its
lower bound $f_{NL}\simeq -9$ still allows for the phantomlike
curvaton to live in a narrow space. From the opposite viewpoint, this
can be another piece of evidence that nature disfavors phantom.

 It is also
worth to note that if the index of equation of state for the
curvaton crosses $-1$ \cite{Feng:2004ad}, then the non-Gaussianity
produced by the curvaton model also crosses $-1$. To realize this
possibility, one usually need more than one curvaton fields
\cite{Xia:2007km}.

Now consider the k-curvaton possibility. If the curvaton has
generalized kinetic terms, then the equilateral non-Gaussianity for
the curvaton is also large. Similar to (\ref{eq:nonlocal}), we have
\begin{equation}
  \label{nonlocal}
  f_{NL}^{\rm (nonlocal)}=\frac{1}{r} f_{NL\chi}^{\rm (nonlocal)}~.
\end{equation}
This amplification can easily produce very large equilateral
non-Gaussianity. Note that $f_{NL}^{\rm (nonlocal)}\sim 1/c_s^2$.
For example, if $1/c_s^2\simeq 5$, and $f_{NL}\simeq 50$, then we
find $f_{NL}^{\rm (nonlocal)}\sim 250$. The experimental
bound $-151<f_{NL}^{\rm (nonlocal)}<253$ (95\% CL). If both large
local and nonlocal $f_{NL}^{\rm (nonlocal)}$ is observed, the
k-curvaton provides a satisfying explanation.

\section{Conclusion and Discussion}

To conclude, in this paper, we have investigated the
 isocurvaton scenario. We found that although the isocurvaton
scenario possesses attractive features such as enhancement of
non-Gaussianity and gravitational waves, the scenario can not be
realized in the slow roll multi-field models. This no-go theorem can
be extended to generalized kinetic terms with assumption $\Xi=0$.
The techniques used in this paper can simplify some calculations in
the mixed curvaton and inflaton scenario, providing an easier
investigation for more general mixed perturbations.

We showed that the no-go result does no harm to the curvaton
scenario. However, the experimental bound on non-Gaussianity,
isocurvature perturbation, and gravitational waves provide a combined
constraint (Eq. (\ref{f5})) on the curvaton model.

We also investigated the phenomenology of phantom and kinetic
curvatons. We showed that the phantom curvaton provides
$f_{NL}\ll-1$, and the k-curvaton provides very large equilateral
non-Gaussianity as well as the local non-Gaussianity.

Finally, let us discuss some possibilities to bypass the no-go
theorem for isocurvaton. The following possibilities are not covered
by the no-go theorem:

\begin{enumerate}
  \item Adding interactions. It is reported from numerical
  calculation that
  interaction can suppress the super horizon perturbations
  \cite{Multamaki:2006tb}. It would be interesting to investigate
  whether similar mechanisms can realize the isocurvaton scenario.

  \item Relaxing the slow roll condition for the isocurvaton field.
  It is challenging to construct fast rolling isocurvaton
  field which can fit the experimental results.

  \item Other form of generalized kinetic terms, including
  separately generalized kinetic terms for inflaton and curvaton
  during inflation, high derivatives like the box term. The
  possibility $\Xi\neq 0$ is also worth investigating.
\end{enumerate}


\section*{Acknowledgment}
This work is supported by grants of NSFC. We thank Xian Gao for
discussions.

\section*{Appendix}
We explain the notation we use, and review some well-known facts in
the cosmological perturbation theory.

In the linear perturbation theory, assuming a flat universe ($K=0$),
and without choosing any gauge, the metric for the scalar
perturbation takes the form
\begin{equation}
  g_{\mu\nu}=\left(
\begin{array}{ccc}
  1+2\phi & -\beta_{,i}\\
  -\beta_{,i}& -a^2((1-2\psi)\delta_{ij}+2E_{,ij})
\end{array}
  \right)~,
\end{equation}
The Newtonian gauge is defined by setting
\begin{equation}
  \beta^{\rm (n)}=0~,~~~E^{\rm (n)}=0~.
\end{equation}

For the $\delta\rho_\varphi=0$ gauge, the equation
$\delta\rho_\varphi=0$ is just one gauge condition, and as in
\cite{NG}, we set the other gauge condition to be $E=0$. In
this notation, the gauge transformation takes the form of Eq.
(\ref{gauge}).

The conserved quantity can be introduced as follows. Assuming that there
is no energy change between $\varphi$ and $\chi$, the local energy
conservation equation for $\varphi$ takes the form
\begin{equation}\label{conserved}
  H-\dot\psi=-\frac{1}{3}\frac{\partial_t(\rho_\varphi+\delta\rho_\varphi)}{\rho_\varphi+\delta\rho_\varphi+p_\varphi+\delta p_\varphi}+{\cal O}\left[\left(\frac{k}{aH}\right)^2\right]~,
\end{equation}
Note that $H$ is a background quantity, and does not change with the
spatial coordinate. If we assume the pressure is a
function of only the energy density, then in the $\delta\rho_\varphi$
gauge, we have $\delta p_\varphi=0$. Thus the RHS of
(\ref{conserved}) is also independent of spacial coordinates. In
order that (\ref{conserved}) holds, $\dot\psi$ should also
independent of spacial coordinates outside the horizon.

As a perturbation variable, $\psi$ should have no zero mode, so does
$\dot\psi$. The only possibility is $\dot\psi=0$. We can
define a conserved quantity
\begin{equation}
  \zeta_\varphi=-\psi \big|_{\delta\rho_\varphi=0}~,
\end{equation}
which is conserved after horizon crossing. This conserved quantity
can be rewritten in the gauge invariant form
\begin{equation}
  \zeta_\varphi=-\psi-H\frac{\delta\rho_\varphi}{\dot\varphi}~.
\end{equation}

From the same reasoning, there is also a gauge invariant conserved
quantity for $\chi$,
\begin{equation}
  \zeta_\chi=-\psi-H\frac{\delta\rho_\chi}{\dot\chi}~.
\end{equation}

Note that these conserved quantities can be defined beyond the
leading order perturbation theory. But we only need the leading
order result for our purpose.

The above proof is under the assumption that in the
$\delta\rho_\varphi=0$ gauge, we have $\delta p_\varphi=0$. This
assumption is obviously true for fluids such as radiation and
matter. It is also worth to note that this assumption is also true
for inflaton after horizon crossing. This is because the above
statement can be rewritten as the adiabatic condition
\begin{equation}
  \dot p_\varphi \delta\rho_\varphi = \dot \rho_\varphi \delta
  p_\varphi~.
\end{equation}
This condition can be checked directly using the Einstein equations
in the $k\ll aH$ limit.

\end{document}